\documentclass{article}

\usepackage{microtype}
\usepackage{graphicx}
\usepackage{subfigure}
\usepackage{booktabs} 

\usepackage{hyperref}

\usepackage[font=small,labelfont=bf]{caption} 
\usepackage{amsmath}        
\usepackage{amsthm}         
\usepackage{amsfonts}       
\usepackage{amssymb}        
\usepackage{wrapfig}        
\usepackage{listings}       
\usepackage{bm}             
\usepackage{xspace}         
\usepackage{bbm}            
\usepackage{makecell}       
\usepackage{float}          

\usepackage[accepted]{icml2019}

\def\compactify{\itemsep=0pt \topsep=0pt \partopsep=0pt \parsep=0pt}
\let\latexusecounter=\usecounter

\usepackage{color}
\usepackage{hyperref}
\definecolor{darkred}{rgb}{0.7,0,0}
\definecolor{darkgreen}{rgb}{0,0.4,0}
\hypersetup{colorlinks=true,
        linkcolor=darkred,
        citecolor=darkgreen}

\usepackage{tikz}
\usetikzlibrary{decorations.pathreplacing}
\makeatletter
\def\underbrace#1{%
   \@ifnextchar_{\tikz@@underbrace{#1}}{\tikz@@underbrace{#1}_{}}}
\def\tikz@@underbrace#1_#2{%
   \tikz[baseline=(a.base)] {\node[inner sep=2] (a) {\(#1\)};
   \draw[line cap=round,decorate,decoration={brace,amplitude=5pt}]
     (a.south east) -- node[pos=0.5,below,inner sep=7pt] {\(\scriptstyle #2\)} (a.south west);}}

\newcommand{\name}{{ABRL}\xspace}

\icmltitlerunning{Real-world Video Adaptation with Reinforcement Learning}

\begin{document}

\twocolumn[
\icmltitle{Real-world Video Adaptation with Reinforcement Learning}




\begin{icmlauthorlist}
\icmlauthor{Hongzi Mao}{mit,fb}
\icmlauthor{Shannon Chen}{fb}
\icmlauthor{Drew Dimmery}{fb}
\icmlauthor{Shaun Singh}{fb}
\icmlauthor{Drew Blaisdell}{fb}
\icmlauthor{Yuandong Tian}{fb}
\icmlauthor{Mohammad Alizadeh}{mit}
\icmlauthor{Eytan Bakshy}{fb}
\end{icmlauthorlist}

\icmlaffiliation{mit}{MIT Computer Science and Artificial Intelligence Laboratory}
\icmlaffiliation{fb}{Facebook}

\icmlcorrespondingauthor{Hongzi Mao}{hongzi@mit.edu}
\icmlcorrespondingauthor{Eytan Bakshy}{ebakshy@fb.com}


\vskip 0.3in
]



\printAffiliationsAndNotice{}  

\begin{abstract}
Client-side video players employ adaptive bitrate (ABR) algorithms to optimize user quality of experience (QoE).We evaluate recently proposed RL-based ABR methods in Facebook's web-based video streaming platform. Real-world ABR contains several challenges that requires customized designs beyond off-the-shelf RL algorithms\,---\,we implement a scalable neural network architecture that supports videos with arbitrary bitrate encodings; we design a training method to cope with the variance resulting from the stochasticity in network conditions; and we leverage constrained Bayesian optimization for reward shaping in order to optimize the conflicting QoE objectives. In a week-long worldwide deployment with more than 30 million video streaming sessions, our RL approach outperforms the existing human-engineered ABR algorithms.
\end{abstract}

\section{Introduction}
\label{s:intro}
The volume of video streaming traffic has been rapidly growing
in recent years~\cite{ciscowhitepaper,sandvine2015}, reaching almost $60\%$
of all the Internet traffic~\cite{sandvine2018global}.
%
%
%
Meanwhile, there has been a steady rise in user demands on video quality\,---\,viewers quickly
leave the video sessions with insufficient quality~\cite{userengagement}.
As a result, content providers are striving to improve the video quality they deliver to the
users~\cite{streamquality}.

Adaptive bitrate (ABR) algorithms are a primary tool 
that content providers use to
optimize video quality subject to bandwidth constraints.
These algorithms run on client-side video players and dynamically choose a bitrate for
each video chunk (e.g., 2-second block), based on network and video observations
such as network throughput measurements and playback buffer occupancy. 
%
%
Their goal is to optimize the video's quality of experience (QoE) by
adapting the video bitrate to the underlying network conditions.
However, designing a strong ABR algorithm with hand-tuned heuristics is difficult, mainly due to hard-to-model
network variations and hard-to-balance conflicting video QoE objectives
(e.g., maximizing bitrate vs. minimizing stalls)~\cite{mpc}. 

Facing these difficulties, recent studies have considered using reinforcement learning (RL)
as a data-driven approach 
to automatically optimize the ABR algorithms~\cite{pensieve}.
%
%
%
RL optimizes its control policy based on the actual performance of past choices, and it is able to
discover policies that outperform algorithms that rely on fixed heuristics or use
inaccurate system models.
For example, as explained in~\citet{pensieve}, RL methods can learn how much playback buffer is necessary to mitigate the risk of
stall in a specific network, based on the network's inherent throughput variability.
%
%
In controlled experiments with a fixed set of videos and network traces, a number of prior work has shown promising results for RL methods~\cite{rl_video_1, rl_video_3}.
However, it remains unknown how the RL-based methods compare to the already deployed heuristic-based ABR methods in large-scale, real-world settings, where generalization and robustness are  
crucial for good performance~\cite{puffer}.
%

In this paper, we present the deployment experience of \emph{\name},
an RL-based ABR module in Facebook's production web-based video platform.
In designing of \name, we found that off-the-shelf RL methods were not sufficient to address
the challenges that we encountered when attempting to deploy RL-based control policies in real-world environments.
To learn high-quality ABR algorithms that surpass the deployed heuristics, we had to
design new components in \name's learning procedure to solve the following challenges.

First, videos in production have different available bitrate encodings, e.g., some videos only
have HD/SD encodings, while other videos have a full spectrum of bitrate encodings. 
However, standard RL approaches use neural networks~\cite{nnbook} that provide fixed outputs both in 
the number of bitrates and the corresponding bitrate levels (e.g., the third output
always corresponds to 720P encoding).
To represent arbitrary bitrate encodings, we design \name's neural network to output a \emph{single}
priority value for each bitrate encoding; and we repeatedly use the same copy of the neural network
for all encodings of a video.
This approach scales to any video \name serves and supports end-to-end RL training (\S\ref{s:rl}).

Second, \name experiences a wide variety of network conditions 
and different video durations 
during training.
This introduces undesirable variance since conventional RL training algorithms cannot tell
whether the observed QoE feedback of two ABR decisions differs due to
disparate network conditions, or due to the quality of the learned ABR policy.
%
To cope with the stochasticity of network conditions, we isolate the rewards on the
actual network trace experienced in a training session, using a recent technique for RL in
environments with stochastic input processes~\cite{variance-reduction}.
This approach separates the contributions of the ABR policy from the overall feedback,
enabling \name to learn robust policies across different deployment
conditions (\S\ref{s:var}).

Third, production ABR requires balancing and co-optimizing multiple objectives together
(e.g., maximize bitrate and minimize stalls). But RL requires a single reward value as the training
feedback. 
Prior work merges the multi-dimensional objectives with a weighted sum~\cite{mpc}.
In practice, since \name's goal is to outperform the existing ABR algorithm in every dimension
of the objective, this does not amount to a specific, pre-defined tradeoff between different objectives.
To determine the weights for different reward components, we 
formulate the multi-objective optimization problem as a constrained optimization problem
(i.e., optimizing one objective subject to bounded degradation along other objectives).
This allows us to use constrained Bayesian
optimization~\cite{constrained_bayes}
to efficiently search for reward weights which best meet top-line objectives (\S\ref{s:reward}).
%
%

Lastly,
for ease of understanding and ensuring safety, 
we translate \name's learned ABR policy
into an interpretable form for deployment. 
Specifically, we realize 
from the policy visualization 
that the learned ABR algorithm exhibits
approximately linear behavior in the observed state of network and buffer occupancy.
Thus, we fit a linear function of network throughput and buffer occupancy
to approximate \name's learned ABR policy (\S\ref{s:translate}).
Such translation degrades the average stall rate by $0.8\%$, but provides 
full interpretability for human engineers.
This allows engineers to understand the policy well enough to verify the learn policy.
%

We run A/B tests that compare \name with the existing ABR algorithms on
Facebook's web-based video streaming platform.
In a week-long worldwide deployment with more than 30 million video streaming
sessions (\S\ref{s:eval}), \name outperforms the heuristic-based ABR policy by $1.6\%$
in average bitrates and reduces stalls by $0.4\%$.
For video sessions with poor network connectivity, in which cases the ABR task
is more challenging, \name provides $5.9\%$ higher bitrate and $2.4\%$ fewer 
stalls.
%
For Facebook, even a small improvement in video QoE is substantial given the scale of its video platform, which consists of millions of hours of video watches per day~\cite{facebookvideohour}.
%
%
In this scale, a fraction of a percent consistent reduction in video buffering is significant; each day, this would save years of video loading time in aggregate.
\section{Background}
\label{s:background}

We provide a review of the basic concepts of adaptive video streaming over HTTP. 
%
%
%
%
Videos are stored as multiple chunks, each of which represents a few
seconds of video playback. Each chunk is encoded at several discrete bitrates, where a
higher bitrate implies a higher resolution and thus a larger chunk size.
The chunks are aligned for seamless transitions across bitrates, i.e., video players can
switch bitrates at any chunk boundary without fetching redundant bits or skipping parts
of the video.

When a client watches a video, the video provider initially sends the client a manifest file that directs
the client to a specific source (e.g., a CDN) hosting the video and lists the available bitrates for the video.
The client then requests video chunks one by one, using an \textit{adaptive bitrate (ABR) algorithm}.
These algorithms use a variety of different inputs (e.g., playback buffer occupancy, throughput measurements, etc.)
to select the bitrate for future chunks. 
As chunks are downloaded, they are stored in the playback buffer on the client. 
Playback of a given chunk
cannot begin until the entire chunk has been downloaded.
In our experiments (\S\ref{s:eval}), we deploy \name on Facebook's web-based
video streaming platform for ease of deployment.

\section{Design}
\label{s:design}
\begin{figure}[t]
\centering
\includegraphics[width=\linewidth]{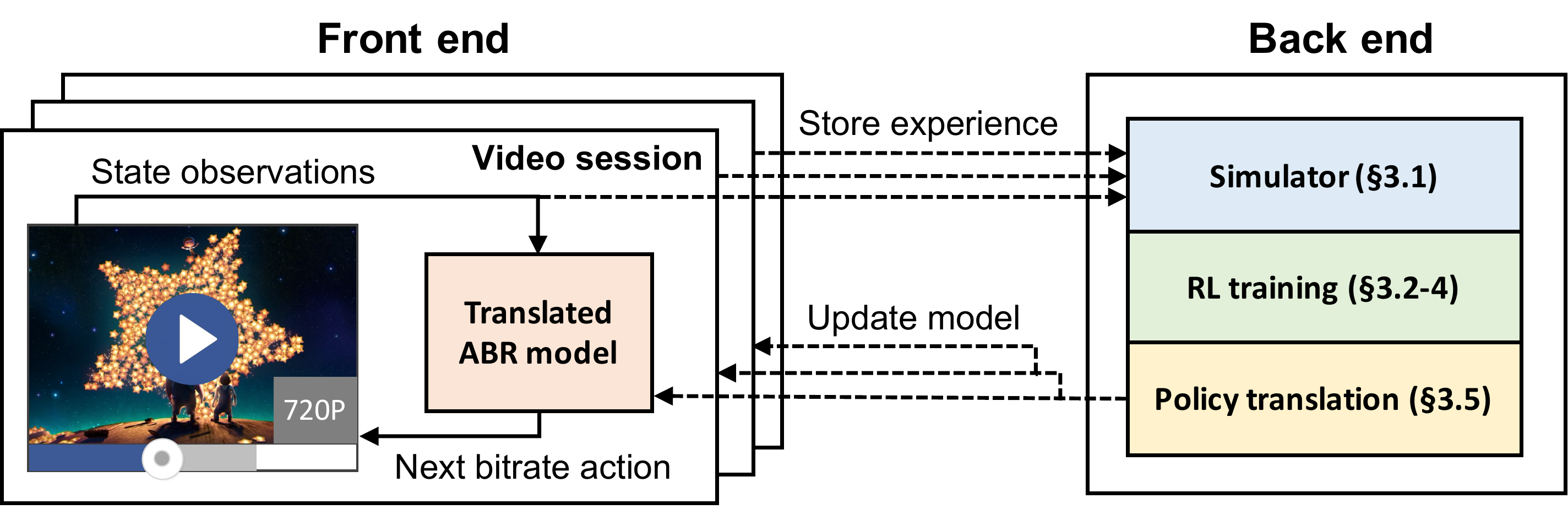}
\caption{Design overview. For each video session in the production experiment, \name collects
the experience of video watch time and the network bandwidth measurements and predictions. It then
simulates the buffer dynamics of the video streaming using these experiences in the backend. After
RL training, \name deploys the translated ABR model to the user front end.}
\label{f:design}
\end{figure}

In this section we describe the design of \name, a system that generates RL-based ABR policies to 
deploy in Facebook's production video platform. We start by describing the simulator that hosts RL 
training in the backend (\S\ref{s:sim}). Next, we explain the RL training framework (\S\ref{s:rl}),
which includes the variance reduction (\S\ref{s:var}) and reward shaping (\S\ref{s:reward}) 
techniques needed for this application. Finally, we describe how \name translates the learned ABR
policy to deploy in the front end (\S\ref{s:translate}). Figure~\ref{f:design} shows an overview.

%
%
\subsection{Simulator}
\label{s:sim}
To train the ABR agent with RL, we first build a simulator that models the playback buffer 
dynamics during video streaming.
The buffer dynamics are governed by the standard ABR procedure described in~\S\ref{s:background}.
Specifically, the simulator maintains an internal representation of the client's playback buffer,
which includes the current size of buffer and the buffer capacity.
The simulator invokes the ABR logic at each video chunk download event, where the ABR logic 
dictates the bitrate decision for the next chunk.
For each chunk download, the simulator determines the download time based on the file size of
the video chunk and the network throughput from the traces.
Since the video is played in real time, the simulator then drains the playback buffer by the
download time of the current chunk representing the video playback during the download. 
If the size of current playback buffer is smaller than the download time, we empty the buffer
and issue a stall event.
Subsequently, the buffer adds the duration of the downloaded chunk into the playback buffer.
In the case where the buffer exceeds the capacity, the simulator ticks the time forward in the
trace without downloading any chunk (i.e., move forward in the bandwidth trace).
The simulated video session terminates at the end of each trace (corresponding to the end of a
watch). During training, \name repeats the simulated video sessions by loading traces
randomly at each time.

The simulator utilizes sampled traces collected from the actual video playback sessions from the frontend.
At each video chunk download event, we log to the backend
a tuple of (1) network bandwidth estimation, (2) bandwidth measurement for the previous chunk
download, (3) the elapsed time of downloading the previous chunk and (4) the file sizes corresponding
to different bitrate encodings of the video chunk.
The bandwidth estimation is an output from a Facebook networking module.
Note that the length of the trace varies naturally across different video sessions due to the difference in the watch time.
In our training, we use more than 100,000 traces from production video streaming sessions.
%
%
\subsection{Reinforcement Learning}
\label{s:rl}
\begin{figure}[t]
\centering
\includegraphics[width=0.95\linewidth]{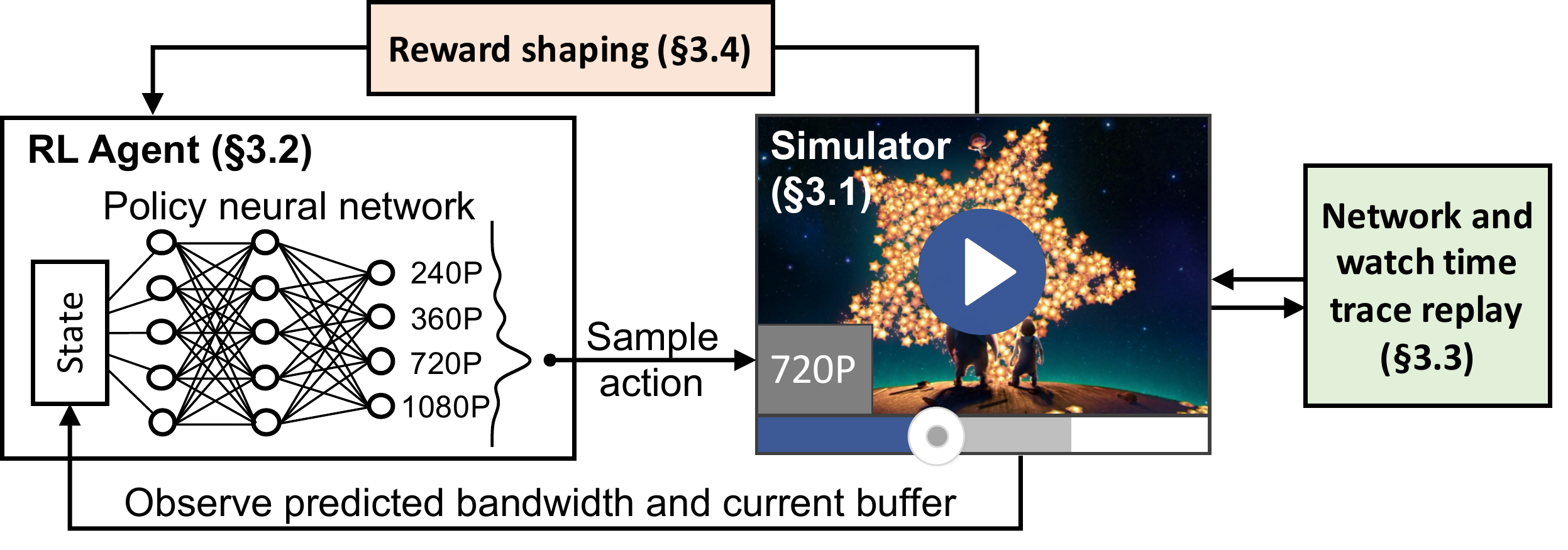}
\caption{Backend RL training framework. \name updates the ABR policy neural network by 
observing the outcome when interacting with a simulator. The simulator uses production traces to simulate the 
video buffer dynamics. 
}
\label{f:rl}
\end{figure}

The training setup shown in Figure~\ref{f:rl} follows the standard RL framework.
In this section, we describe the details of the RL agent and the policy gradient training algorithm.
In particular, we explain the challenges we encountered to motivate the variance reduction (\S\ref{s:var}) and the reward shaping (\S\ref{s:reward}) techniques.

\textbf{RL setup.}
Upon downloading each video chunk at each step $t$, the RL agent observes the \emph{state} 
$s_t = (x_t, o_t, \vec{n}_t)$, where $x_t$ is the bandwidth prediction for the next chunk,
$o_t$ is the current buffer occupancy and $\vec{n}_t$ is a vector of the file sizes for the next video chunk.
As a feedback for the bitrate \emph{action} $a_t$, the agent receives a \emph{reward} $r_t$
constructed as a weighted combination of selected bitrate $b_t$ and stall time of the past chunk $d_t$:
\begin{align}
	r_t = w_b b_t^{v_b} - w_d d_t^{v_d} + w_c [\mathbbm{1} (d_t > 0)], \label{eq:reward}
\end{align}
where $\mathbbm{1}(\cdot)$ is an indicator function counting the number of stalls,
and $w_b, w_d, w_c, v_b, v_d$ are the tuning weights for the reward.
Notice that these weights cannot be predetermined, 
because the goal of RL-based ABR is to outperform the existing ABR
algorithm in every dimension of the metric (i.e., higher bitrate, less stall time and less stall count),
which does not amount to a quantitative objective.
In \S\ref{s:reward} we describe how we use Bayesian optimization to shape the weights for optimizing the
multi-dimensional objective.

\begin{figure}[t]
\centering
\includegraphics[width=0.95\linewidth]{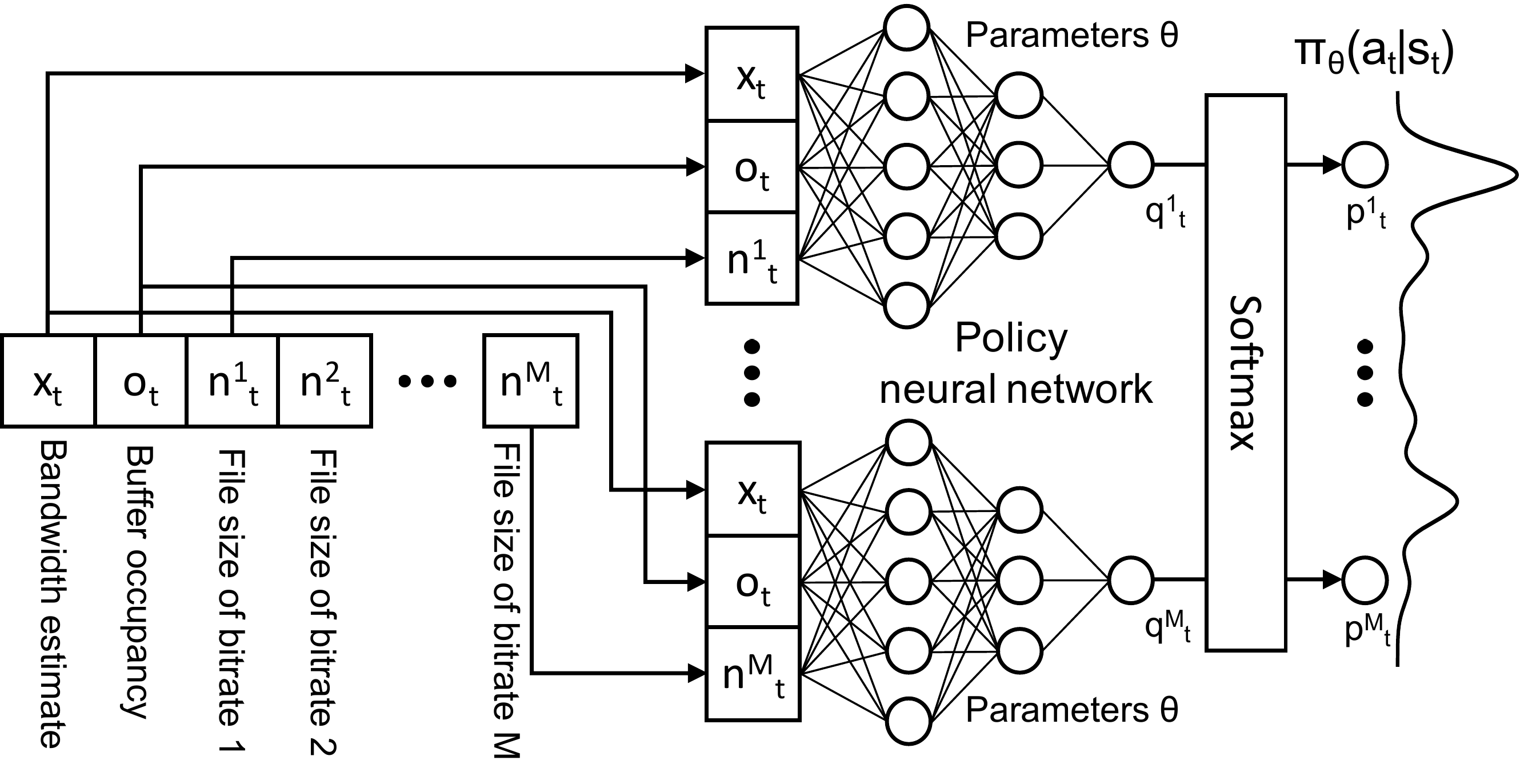}
\caption{Policy network architecture. For each bitrate, the input is fed to a copy of the \emph{same}
policy neural network. We then apply a (parameter-free) softmax operator to compute the probability distribution
of the next bitrate. This architecture can scale to arbitrary number of bitrate encodings.
}
\label{f:policy}
\end{figure}

\textbf{Policy.} 
As shown in Figure~\ref{f:policy}, the agent samples the next bitrate action $a_t$ based on
its parametrized policy: 
$\pi_\theta(a_t | s_t) \to [0, 1]$.
%
%
In practice, since the number of bitrate encodings (thus the length of $\vec{n}_t$) varies across
different videos~\cite{many_encodings}, 
we architect the policy network to take an arbitrary number of file sizes as input.
Specifically, for each bitrate, the input to the policy network consists of the predicted bandwidth and buffer occupancy, concatenated with the corresponding file size.
We then copy the \emph{same} neural network for each of the bitrate encodings (e.g., the neural
networks shown in Figure~\ref{f:policy} share the same weights $\theta$).
Each copy of the policy network outputs a ``priority'' value $q^i_t$ for selecting the corresponding
bitrate $i$.
Afterwards, we use a softmax~\cite{bishop} operation to map these priority values into a probability distribution $p^i_t$ over each bitrate: $p^i = \exp(q^i_t) /  \sum_{i=1}^M[\exp(q^i_t)]$.
Importantly, the whole policy network architecture is end-to-end differentiable and can be trained
with the policy gradient algorithms~\cite{pg}.

\textbf{Training.}
%
%
We use the policy gradient method~\cite{pg, rlbook, elf} to update the policy neural network parameters
in order to optimize for the objective.
%
%
Consider a simulated video streaming session of length $T$, where the agent collects (state, action,
reward) experiences, i.e., $(s_t, a_t, r_t)$ at each step $t$.
The policy gradient method updates the policy parameter $\theta$ using the estimated gradient of the cumulative reward:
\begin{align}
	\theta \leftarrow \theta + \alpha \sum_{t=1}^{T} \nabla_\theta \log \pi_\theta(s_t, a_t) \left( \sum_{t' = t}^{T} r_{t'} - b_t \right), \label{eq:pg}
\end{align}
where $\alpha$ is the learning rate and $b_k$ is a {\em baseline} for reducing the variance of the policy gradient~\cite{baseline}.
%
%

Notice that the estimation of the advantage over the average case relies on the accurate estimation of
the average. For this problem, the standard baselines, such as the time-based baseline~\cite{greensmith, williams_baseline} or value function~\cite{a3c}, suffer from large variance due to the stochasticity in the traces~\cite{variance-reduction}. We further describe the details of this variance in \S\ref{s:var} and our approach to reducing it.

%
%
\subsection{Variance Reduction}
\label{s:var}
\begin{figure}[t]
\centering
\includegraphics[width=0.85\linewidth]{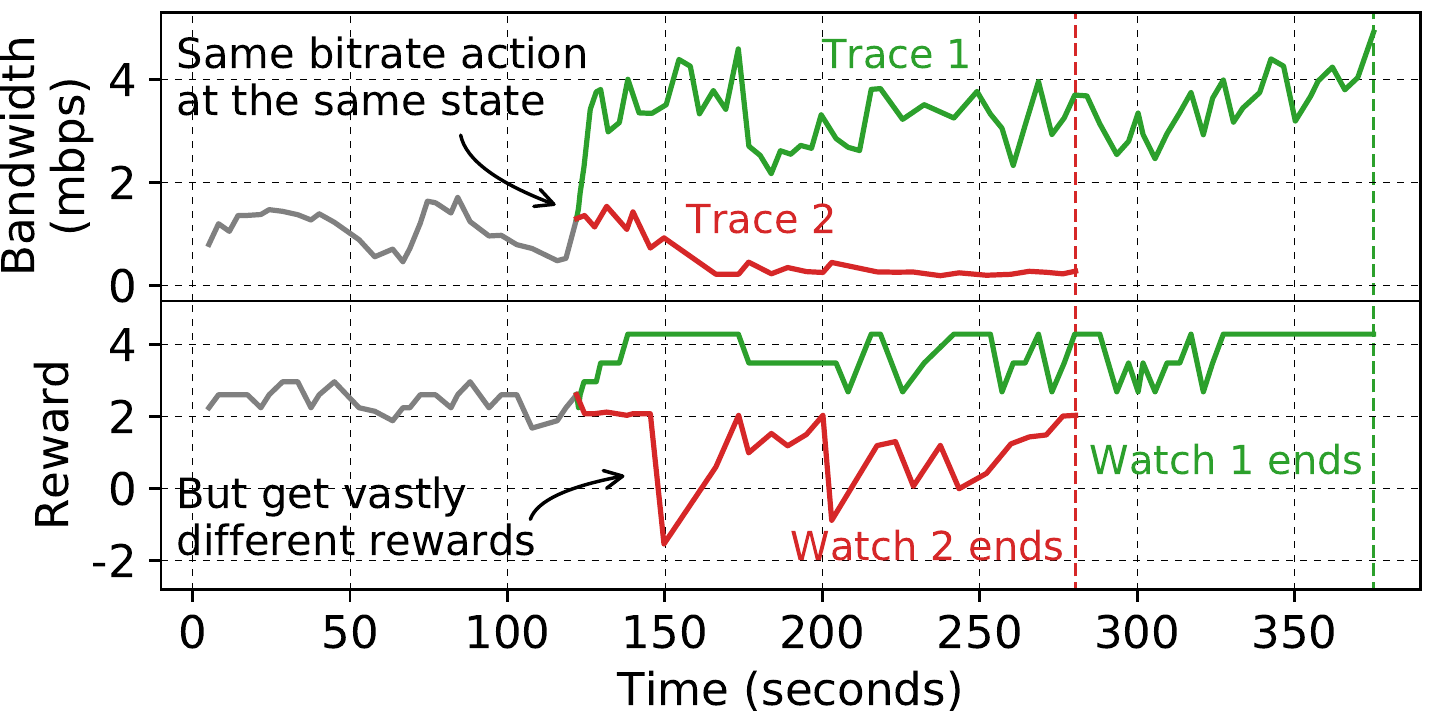}
\caption{Illustrative example of how the difference in the traces of network
bandwidth and video watch time creates significant variance for the reward feedback.
}
\label{f:reward}
\end{figure}

\name's RL training on the simulator is powered by a large number
of network traces collected from the front end video platform (\S\ref{s:sim}).
During training, \name must experience a wide variety of network conditions
and video watches in order to generalize its ABR policy well.
However, this creates a challenge for training: different traces contain very
different network bandwidth and video duration, which significantly affects
the total reward observed by the RL agent.
Consider an illustrative example shown in Figure~\ref{f:reward}, where
we use a fixed buffer-based ABR policy~\cite{bufferabr} to make the
bitrate action at time $\tau$.
Even for this fixed policy, if the future trace happens to contain large 
bandwidth (e.g., Trace 1), the reward feedback will naturally be large, since 
the network can support high bitrate without stalls.
In contrast, if the future network condition becomes poor (e.g., Trace 2), the 
reward will likely be lower than average.
More importantly, the video duration determines the possible length of ABR 
interactions, which dictates the \emph{total} reward the RL agent can receive for
training (e.g., the longer watch time in Trace 1 leads to larger total reward).
The key problem is that the difference across the traces is independent with
the bitrate action at time $\tau$\,---\,e.g., the future bandwidth might
fluctuate due to the inherent stochasticity in the network; or a user might stop watching a video regardless of the quality.
As a result, this creates large variance in the reward feedback used for estimating the policy gradient in Equation~\eqref{eq:pg}.

To solve this problem, we adopt a recently proposed technique for handling
an exogenous, stochastic process in the environment when training RL
agents~\cite{variance-reduction}. 
The key idea is to modify the baseline in Equation~\eqref{eq:pg} to
an ``input-dependent'' one that takes the input process (e.g., the trace 
in this problem) into account explicitly.
%
%
In particular, for this problem, we implement the input-dependent baseline
by loading the \emph{same} trace (i.e., the same time-series for network 
bandwidth and the same video watch time) multiple times and computing the
average total reward at each time step among these video sessions.
Essentially, this uses the time-based baseline~\cite{greensmith}
for Equation~\eqref{eq:pg} but computes the average return conditional
on the specific instantiation of a trace.
During training, we repeat this procedure for a large number of randomly-sampled 
network traces.
As a result, this approach entirely removes the variance caused by the
difference in future network condition or the video duration.
Since the difference in the reward feedback is only due to the difference
in the actions, this enables the RL agent to assess the quality of different 
actions much more accurately. 
In Figure~\ref{f:reward_var}, we show how this approach helps improve the training
performance.
%
%
\subsection{Reward Shaping with Bayesian Optimization}
\label{s:reward}
%
The goal of \name is to outperform the existing ABR policy according to multiple team-wide objectives (i.e., increasing the video quality while reducing the stall time).
Recall that the reward weights in Equation~\eqref{eq:reward} dictates the performance of \name's
learned ABR policy in each of the objective dimensions.
These objectives have an inherent trade-off: optimizing one dimension (by tuning up the corresponding
reward weight) diminishes the performance in another dimension (e.g., high video quality increases the 
risk of stalls).

To determine the proper combination of the reward weights, we treat \name's RL training module 
(\S\ref{s:rl}, \S\ref{s:var}) as a black box function $f(\vec{w})\rightarrow(q, l)$ 
that maps the reward weights $\vec{w} \triangleq (w_b, w_d, w_c, v_b, v_d)$ 
from Equation~\eqref{eq:reward} to a noisy estimate of the 
average video quality $q$ and stall rate $l$ in unseen test video sessions.

Then, we use Bayesian optimization~\cite{ human_out_of_bayes} to efficiently search for the
weight combinations that leads to better $(q, l)$, with only a few invocations of the RL training module.  This procedure of tuning the weights in the reward function is a realization of reward
shaping~\cite{reward-shaping}.
We formulate the multi-dimensional optimization problem as a constrained optimization problem:
\begin{align}
	\text{argmax}_{\vec{w}}~ 
	q(\vec{w}),~
	\text{subject to}~ 
	\frac{l(\vec{w})}{l_s} \leq C
\end{align}

Where $q(\vec{w})$ and $l(\vec{w})$ are the quality and stall rate evaluated at $\vec{w}$, $l_s$ is the stall rate of the existing policy (non-RL based) used in production at Facebook, and $C$ is some constraint value. 

Notice that the function $q(\cdot)$ and $l(\cdot)$ are can only be observed
by running the RL training module---a computationally intensive procedure.  We solve this constrained optimization problem with Bayesian optimization.  Bayesian optimization uses a Gaussian process (GP) ~\cite{gaussian_process} surrogate model to approximate the results of the RL training procedure using a limited number of training runs.  Gaussian processes are flexible non-parametric Bayesian models representing a posterior distribution over possible smooth functions compatible with the data.  We find that GPs are excellent models of the output of the RL training module, as small changes to the reward function will result in small changes in the overall outcomes.  Furthermore, GPs are known to produce good estimates of uncertainty.

Using Bayesian optimization, we start from an initial set of $M$ design points $\{\vec{w_i}\}_{i=1}^M$, and iteratively test new points on the RL module according to an acquisition function that navigates the explore/exploit tradeoff based on a surrogate model (most commonly a GP).

A popular acquisition function for Bayesian optimization is expected improvement (EI)
(see \citet[\S4.1]{bayesopt_tutorial}).
The basic version of EI simply computes the expected value of improvement at each point relative to the best observed point: 
$\alpha_{EI}(\vec{x}\,|\, f^*) = \mathbb{E}_{y \sim g(\vec{x}\,| \mathcal{D})} [\max(0, f(y) - f^*)]$,
where $\mathcal{D} \triangleq \{\vec{w}_i, q(\vec{w}_i)\}_{i=1}^N$ represents $N$ runs of data points,
$f^*$ is the current best observed value and $g(\vec{x}\,|\mathcal{D})$ denotes the
the posterior distribution of $f$ value from the surrogate.

We use a variant of EI---Noisy Expected Improvement (NEI)---which supports optimization of noisy, constrained function evaluations~\citep{constrained_bayes}.  While EI and its constrained variants (e.g., \citep{constrained_bayes}), are designed to optimize deterministic functions (which have a known best feasible values), NEI integrates over the uncertainty in which observed points are best, and weights the value of each point by the probability of feasibility.

NEI naturally fits the structure of the optimization task, since the training procedure is stochastic (e.g., it depends on the random seed). We therefore evaluate the \name RL training module with a given $\vec{w}$ multiple times and compute its standard error, which are then passed into the NEI algorithm.  NEI supports batch updating, allowing us to evaluate multiple reward parameterizations in parallel.

\subsection{Policy Translation}
\label{s:translate}
In practice, most video players execute the ABR algorithms in the front end to avoid the extra latency 
connecting to the back end~\cite{dashstudy, dashstandard, youtubeplayer, bufferabr}.
Therefore, we need to deploy the learned ABR policy to the users directly\,---\,i.e., the design of an
ABR server in the back end hosting the requests from all users is not ideal~\cite{pensieve}.
To massively deploy, we make use of the web-based video platform at Facebook, where the front end
service (if uncached) fetches the most up-to-date video player (including the ABR policy) from
the back end server at the beginning of a video streaming session.
For ease of understanding and maintenance in deployment, we translate the neural network ABR
policy to an interpretable form.
In particular, we found that the learned ABR policies approximately exhibit a linear structure\,---\,the
bitrate decision boundaries
are approximately linear and the distances between the boundaries are constant in part of the decision space.
As a result, we approximate the learned ABR policy with a deterministic linear fitting function.
Specifically, we first randomly pick $N$ tuples of bandwidth prediction $x$ and buffer occupancy $o$
(see the inputs in Figure~\ref{f:policy}). Then, for each tuple values $(x, o)$ and for each of
the $M$ equally spaced bitrates with file sizes $n^1, n^2, \cdots, n^M$, we invoke the policy network
to compute the probability of selecting the corresponding bitrate:
$\pi(a^1 | x, o, n^1), \pi(a^2 | x, o, n^2), \cdots, \pi(a^M | x, o, n^M)$. 
Next, we determine the ``intended'' bitrate
using a weighted sum: $\bar{n} = \sum_{i=1}^M n^i \pi(a^i |x, o, n^i)$. This serves as the target bitrate for
the output of the linear fitting function.
Finally, we use three parameters $a, b, $ and $c$, to fit a linear model of bandwidth prediction and buffer
occupancy, which minimizes the mean squared error over all $N$ points:
\begin{align}
	\sum_{i=1}^N \big| a x_i + b o_i + c - \bar{n}_i \big|^2. \label{eq:lin}
\end{align}
Here, we use the standard least square estimator for the model fitting, which is the optimal
unbiased linear estimator~\cite{ols-blue}.
At inference time, the front end video player uses the fitted linear model to determine the
intended bitrate and then selects the maximum available bitrate that is below the intended bitrate. 
%

%
Translating the neural network ABR policy provides interpretability for human engineers but
it is also a compromise in terms of ABR performance (\S\ref{s:analysis} empirically evaluates this trade-off). 
Also, adding more contextual based features would likely require a non-linear policy encoded directly
in a neural network (\S\ref{s:discussion}).
It is worth noting that directly using RL to train a linear policy is a natural choice.
%
%
However, to our surprise, training \name with a linear policy function leads to worse ABR performance
than the existing heuristics.
We hypothesize this is because policy gradient with a weak function approximator such as
a linear one has difficulty converging to the optimal, even though the optimal policy can be 
simple~\cite{delusional_q_learn, pg_div, val_iter_div, deadly_triad}.

\section{Experiments}
\label{s:eval}
We evaluate \name with Facebook's web-based production video platform. 
Our experiments answer the following questions:
(1) Does \name provide gains in performance over the existing heuristic-based production ABR algorithm? 
(2) How are different subgroups affected by the \name-based policy?
(3) What ABR policy does \name learn?
(4) During training, how do different design components affect the learning procedure?


\subsection{Overall live performance}
\label{s:perf}
\begin{figure}[t]
\centering
\subfigure[Video quality]{\label{f:overall_quality}
\includegraphics[width=0.48\linewidth]{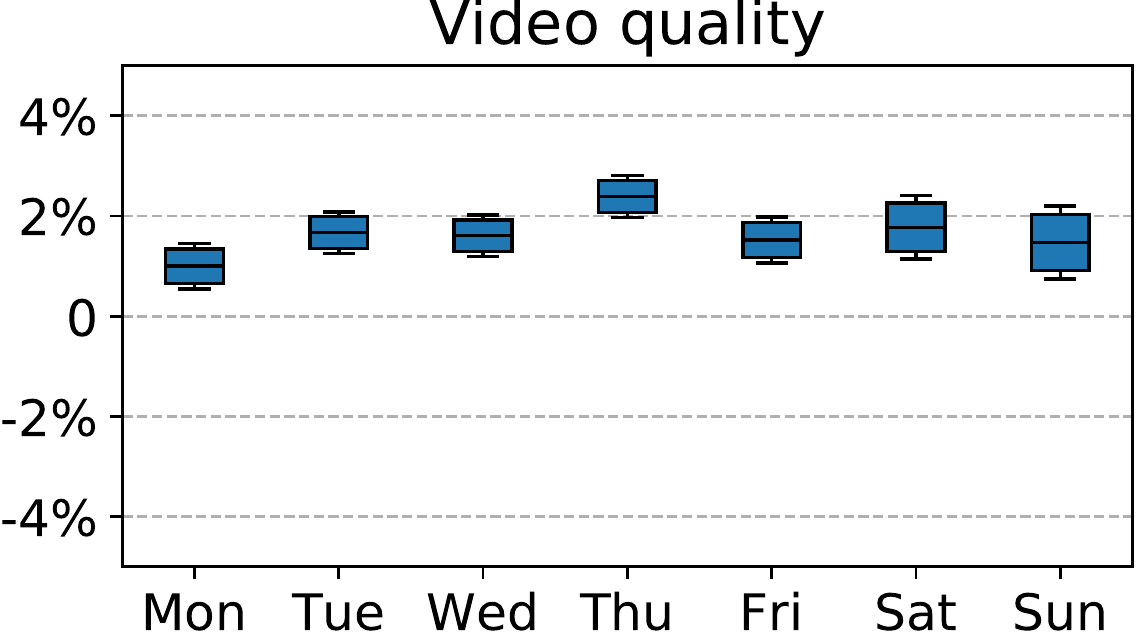}}
\subfigure[Stall rate]{\label{f:overall_stall}
\includegraphics[width=0.48\linewidth]{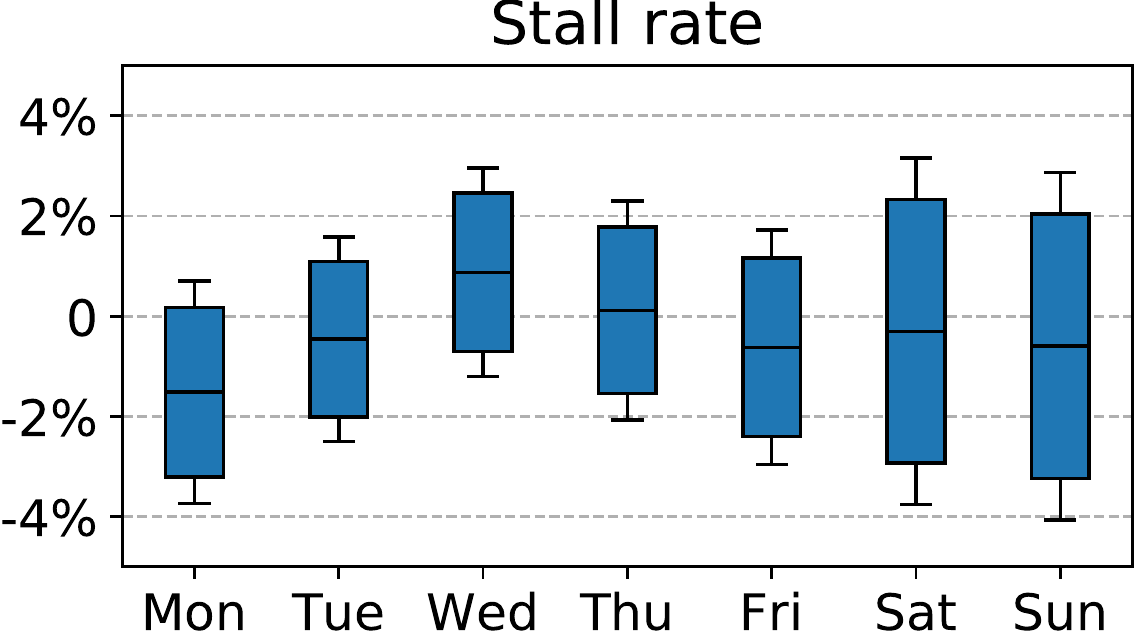}}
%
%
\vspace{-0.3cm}
\caption{A week-long performance comparison with production ABR policy.
The comparison is sampled from over 30 million video streaming session.
The box spans $95\%$ confidence intervals and the bars spans $99\%$
confidence intervals.}
\label{f:overall_perf}
\end{figure}

In a week-long deployment on Facebook's production video platform, we compare the performance 
of \name's translated ABR policy (\S\ref{s:translate}) with that of the existing heuristic-based ABR
algorithm. 
The experiment includes over 30 million worldwide video playback sessions.
Figure~\ref{f:overall_perf} shows the relative improvement of \name in terms of video quality and stall rate.

Overall, \name achieves a $1.6\%$ increase in average bitrate and a $0.4\%$ decrease in stall rate.
Most notably, \name consistently selects higher bitrate through the whole week ($99\%$ confidence intervals all positive). 
However, choosing higher bitrates does \emph{not} sacrifice stall rate\,---\,
\name rivals or outperforms the default scheme on the average stall rate every day, even on Thursday
when gains in video quality are highest.
This shows \name uses the output from the bandwidth prediction module better than the
fine-tuned heuristic. By directly interacting with the observed data, \name learns quantitatively
how conservative or aggressive the ABR should be with different predicted bandwidths.
As a result, this also leads to a $0.2\%$ improvement in the end-user video watch time.

These improvement numbers may look modest compared to the those reported by recent 
academic papers~\cite{bufferabr, bola, mpc, pensieve}. This is mostly because we only experiment 
with web based videos, which primarily consist of well-connected desktop or laptop traffic,
different from the prior schemes that mostly concern cellular and unstable networks.
Nonetheless, any non-zero improvement is significant given the massive volume of Facebook videos.
In the following, we profile the
performance gain at a more granular level.
%
\subsection{Detailed Analysis of RL Pipeline}
\label{s:analysis}

\begin{figure}[t]
\centering
\includegraphics[width=0.75\linewidth]{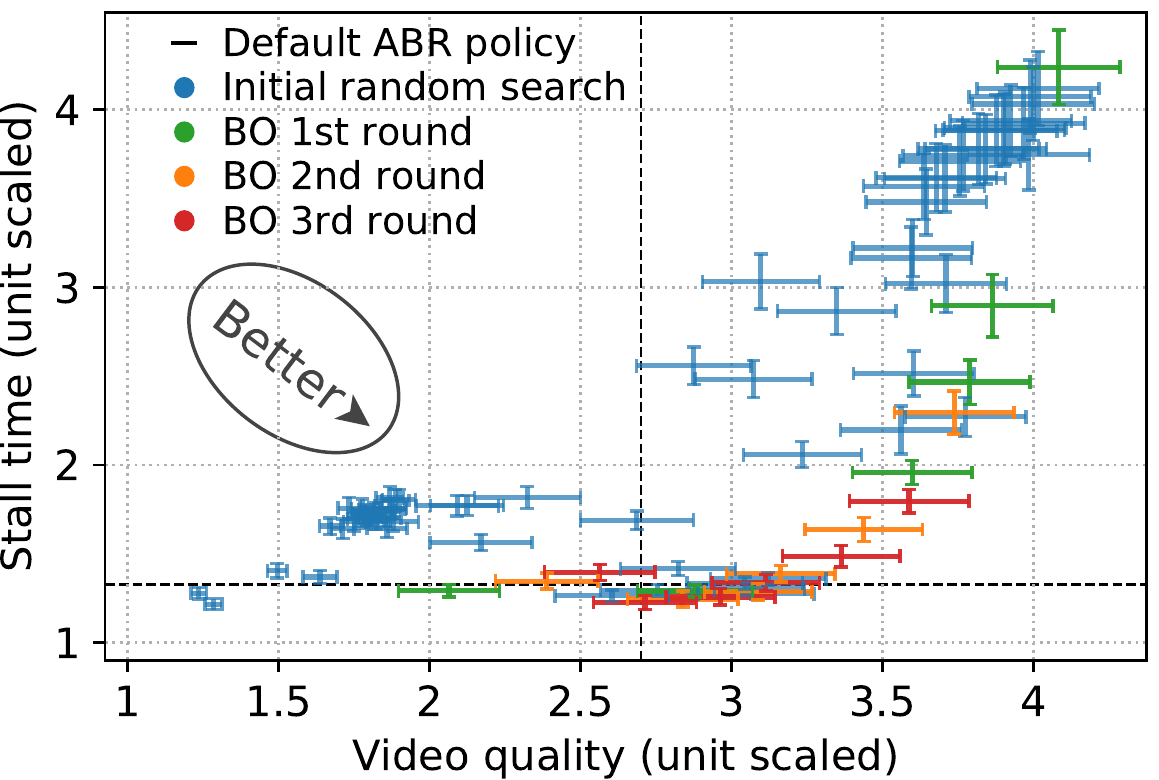}
\caption{Reward shaping via Bayesian optimization using the \name simulator. The initial round has 64 random initial parameters. Successive batches of Bayesian optimization converge to optimal weightings that improve video quality while reducing stall rate. The performance is tested on held out network traces.}
\label{f:reward_shaping}
\vspace{-1em}
\end{figure}

\textbf{Reward shaping.}
To optimize the multi-dimensional objective, we use a Bayesian Optimization approach for reward
shaping (\S\ref{s:reward}).
The goal is to tune the weights in the reward function in order to train a policy that operates
on the Pareto frontier of video quality and stall (and, ideally, outperform the existing policy
in both dimensions).
Figure~\ref{f:reward_shaping} shows the performance from different reward weights during the reward shaping procedure. 
At each iteration, we set the reward weights using the output from the Bayesian optimization module,
and treat \name's RL module as a black box, in which the policy is trained until convergence according to the chosen reward weights.
The Bayesian optimization module then observes the testing outcomes (both video quality and stall) and
sets the search criteria for the next iteration to be ``expected improvement in video quality such that stall time
degrades no more than $5\%$''.
As shown, within three iterations, \name is able to hone in on the empirical Pareto frontier.
In this search space, there are many more weight configurations that lead to better video quality
(i.e., right of the dashed line) than the configurations leading to fewer stalls (i.e., lower than the
dashed line).
Compared to the existing ABR scheme, \name finds a few candidate reward weights that lead to better
ABR policy both in terms of video quality and stalls (i.e., lower and to the right of the existing policy). 
For the production experiment in \S\ref{s:perf}, we deploy the policies within the region that 
shows the largest improvement in stall.
After this search procedure, engineers on the video team can pick policies based on different deployment objectives as well.
%

\begin{figure}[t]
\centering
\includegraphics[width=0.7\linewidth]{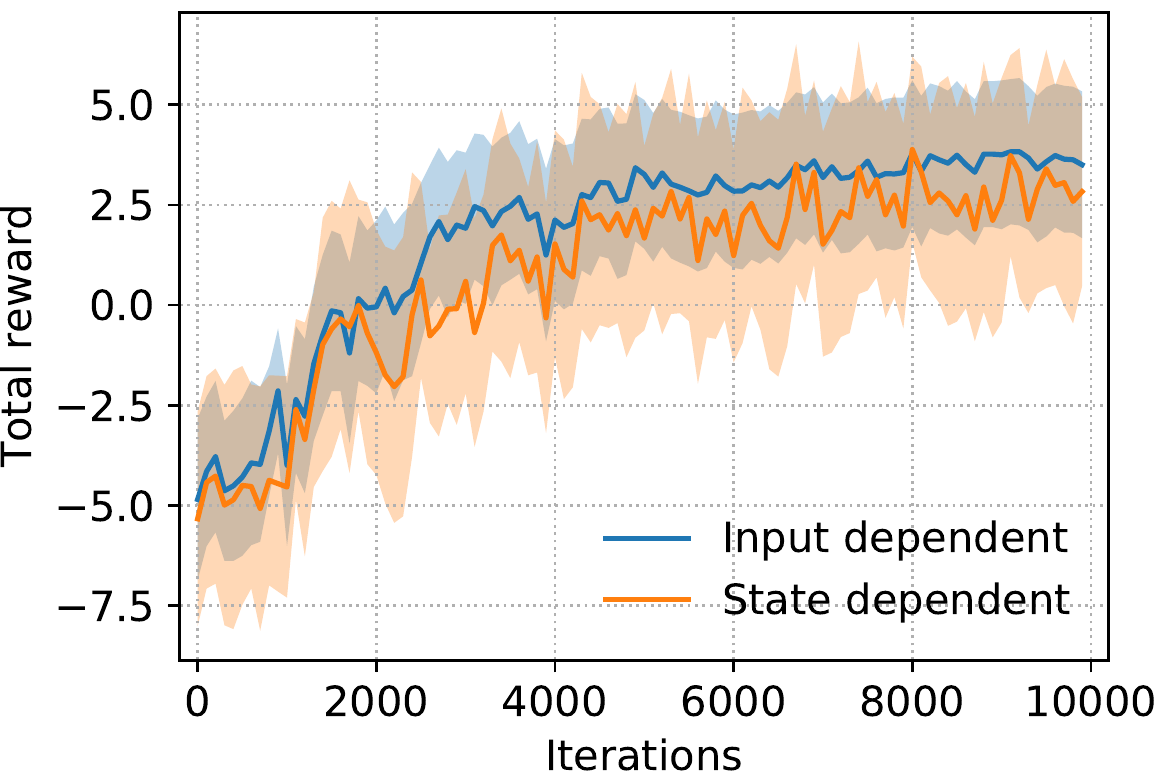}
\vspace{-0.3cm}
\caption{Improvements learning performance due to variance reduction. The network condition and watch time in different traces introduces variance in the policy gradient estimation. The input-dependent baseline helps reduce such variance and improve training performance. Shaded area spans $\pm$ std.}
\label{f:reward_var}
\end{figure}

\textbf{Variance reduction.} 
To reduce the variance introduced by the network and the watch time across different the traces,
we compute the baseline for policy gradient by averaging over the cumulative rewards from
the same trace (in all the parallel rollouts) at each iteration, effectively achieving the
input-dependent baseline (\S\ref{s:var}).
For comparison, we also train an agent with the regular state-dependent baseline (i.e., output
from a value function that only takes the state observation as input).
Figure~\ref{f:reward_var} evaluates the impact of variance reduction by comparing the learning
curve trained with the input-dependent baseline to that with the state-dependent baseline. 
As shown, the agent with the input-dependent baseline achieves about $12\%$ higher eventual total reward
(i.e., the direct objective of RL training).
Moreover, we find that the agent with input-dependent baseline converges faster in terms of the entropy
of the policy, which is also indicated by the narrower shaded area in Figure~\ref{f:reward_var}.
At each point in the learning curve, the standard deviation of rewards is around half as large under the input-dependent baseline.
This is expected because of the large variance in the policy gradient estimation given the
uncertainties in the trace. Fixing the trace at each training iteration removes the variance introduced
by the external input process, making the training significantly more stable.

\begin{figure}[t]
\centering
\subfigure[Video quality]{\label{f:video_quality_compromise}
\includegraphics[width=0.48\linewidth]{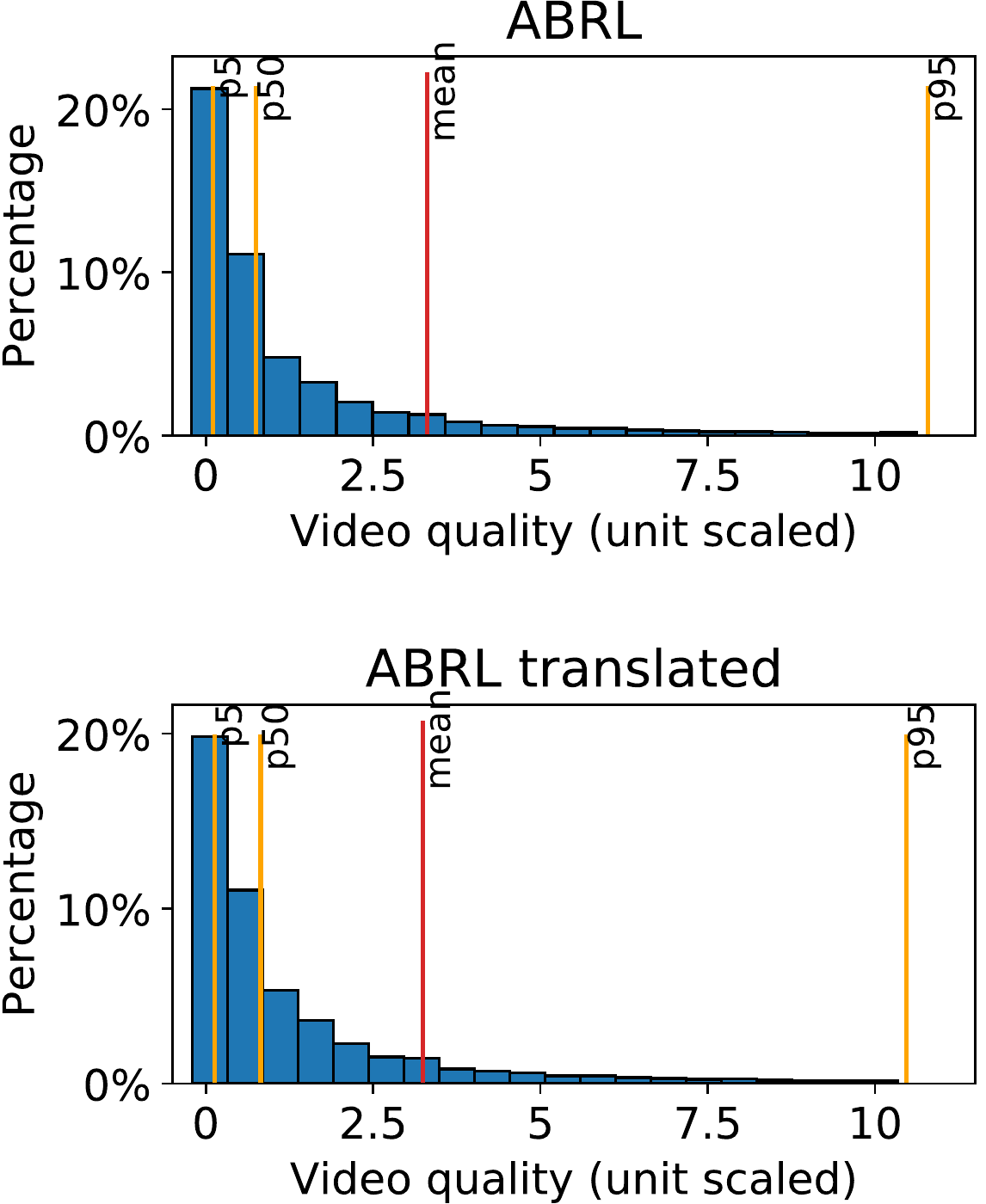}}
\subfigure[Stall]{\label{f:stall_compromise}
\includegraphics[width=0.48\linewidth]{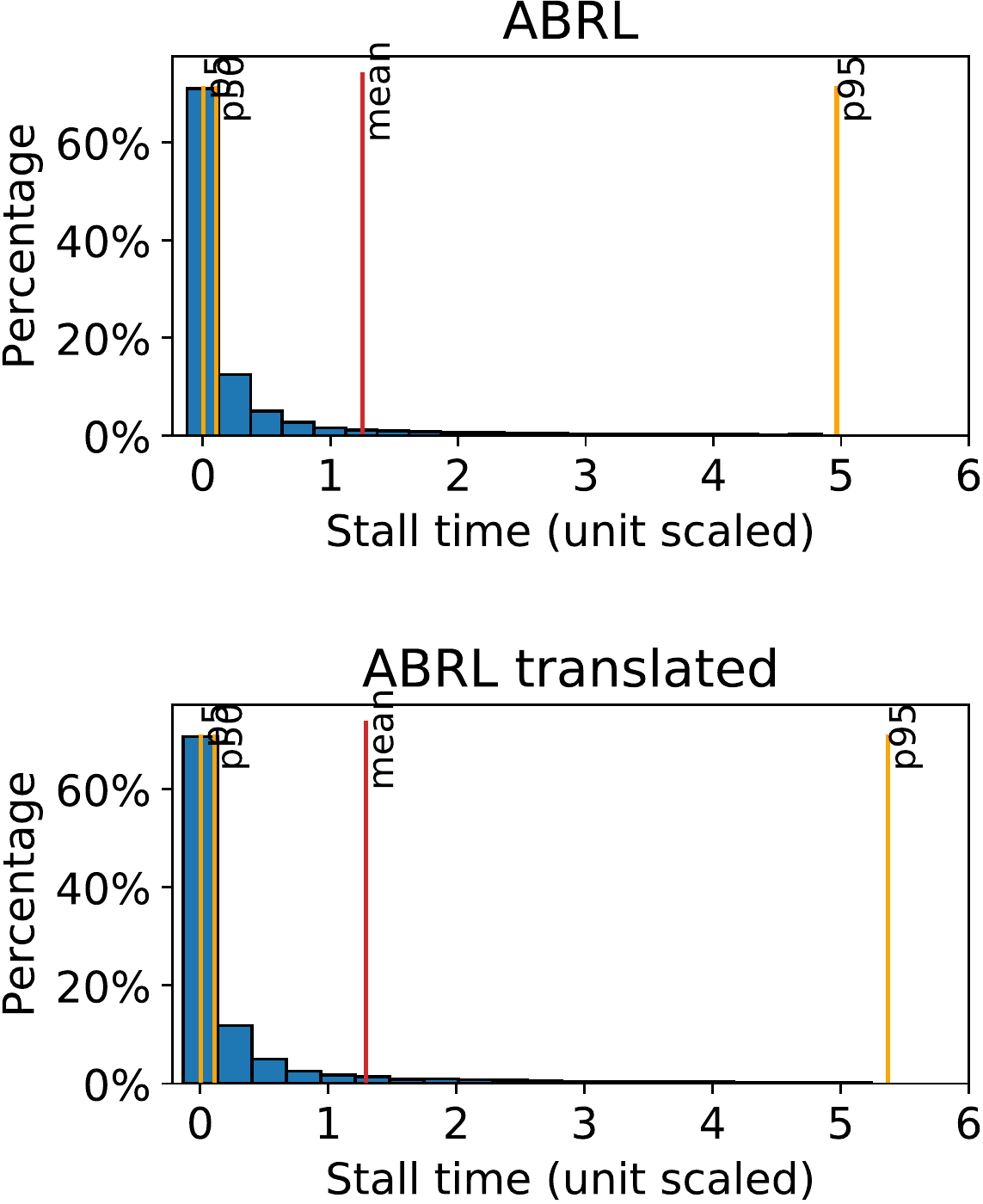}}
\vspace{-0.3cm}
\caption{Performance comparison of \name and its linear approximated variant. The agents are tested with
unseen traces in simulation. Translating the policy degrades the average performance by $0.8\%$ in stall and
$0.6\%$ in quality.
}
\label{f:perf_compromise}
\end{figure}

\textbf{Trade-off of performance for interpretability.}
Figure~\ref{f:perf_compromise} shows how the testing performance of video quality and stall in simulation
differ between \name's original neural network policy and the translated policy (\S\ref{s:translate}). 
Most noticeably, making the ABR policy linear and interpretable incurs a $0.8\%$
and $8.9\%$ degradation in the mean and 95th percentile of stall rate.
This accounts for the tradeoff to make the learned ABR policy fully interpretable.
Also, we tried to 
train a linear policy directly from scratch (by removing hidden layers in the neural network and 
removing all the non-linear transformations). However, the performance of the directly learned linear policy does not outperform the existing baseline. 
This in part is because
over-parametrization in the policy network helps learn a more robust policy~\cite{delusional_q_learn, val_iter_div}.

\textbf{Subgroup analysis.}
To better understand how \name outperforms the existing ABR scheme, we breakdown
the performance gain in different network conditions and we visualize the ABR policy
learned by \name.

\begin{figure}[t]
\centering
\subfigure[Video quality breakdown]{\label{f:video_quality_breakdown}
\includegraphics[width=0.48\linewidth]{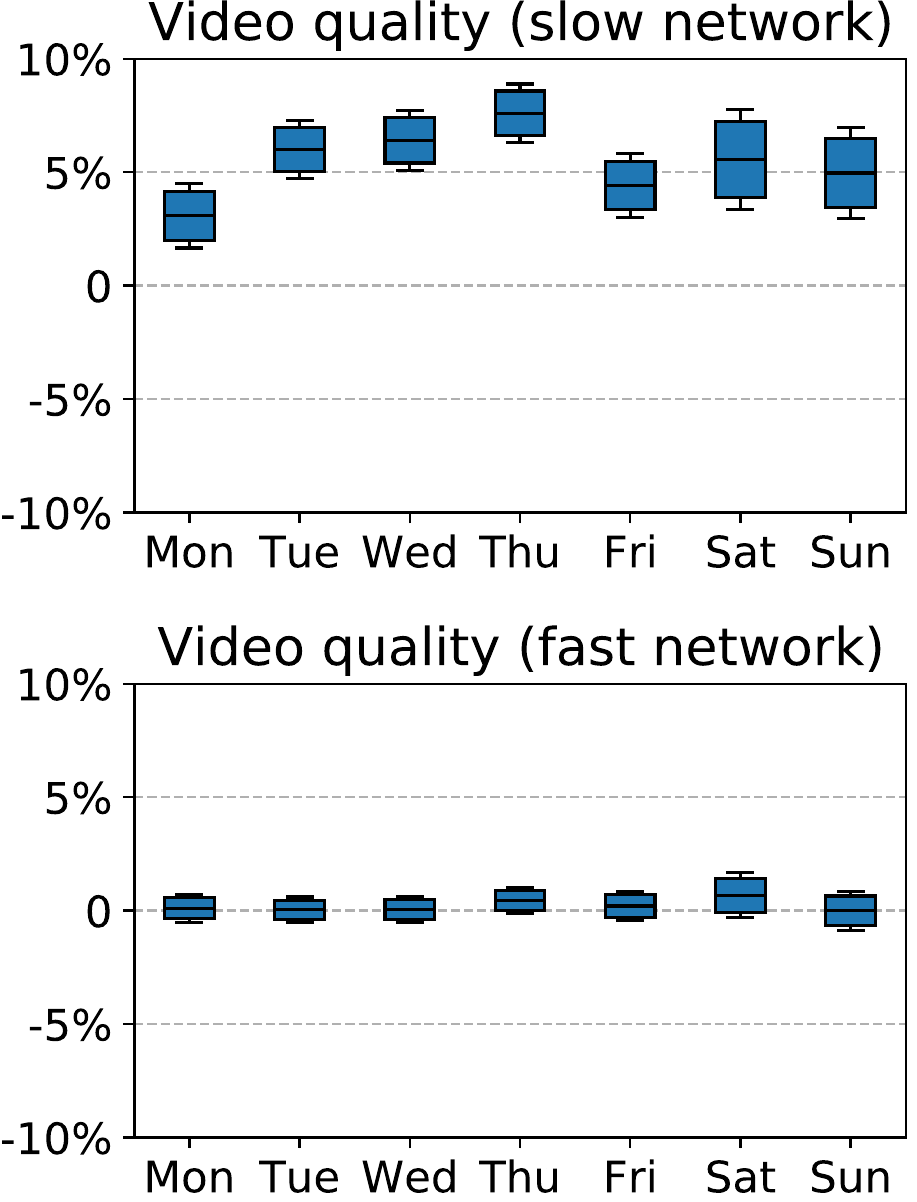}}
\subfigure[Stall rate breakdown]{\label{f:stall_breakdown}
\includegraphics[width=0.48\linewidth]{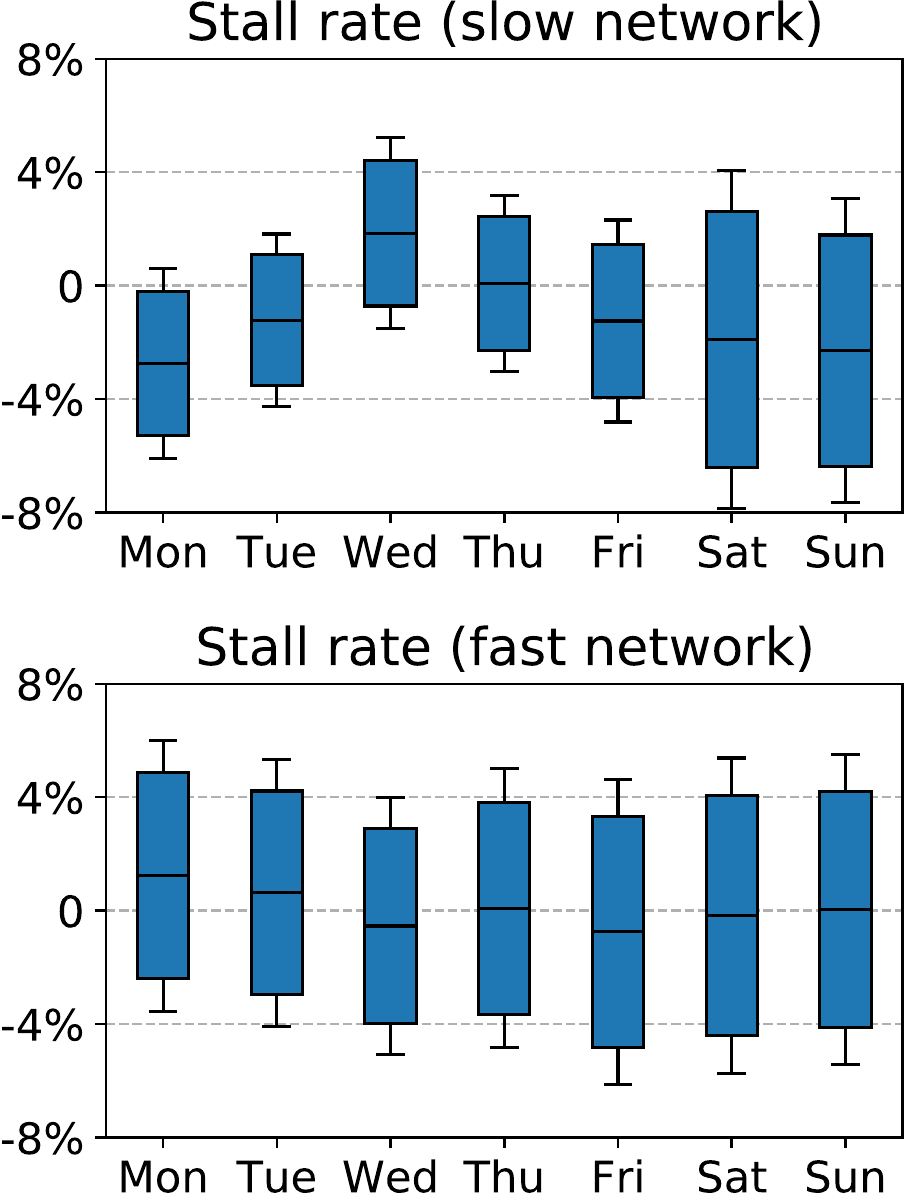}}
\caption{Breakdown the performance comparison with different network
quality for the live experiment. ``slow network'' corresponds to $<500$ kbps measured network bandwidth,
and ``fast network'' corresponds to $>10$ mbps bandwidth. 
The box spans $95\%$ confidence intervals and the bars spans $99\%$
confidence intervals.}
\vspace{-1em}
\label{f:perf_breakdown}
\end{figure}
In Figure~\ref{f:perf_breakdown}, we categorize the video sessions based on the average measured
network bandwidths.
As shown, \name overall achieves a higher bitrate while maintaining fewer stalls in both fast and
slow networks. 
Moreover, \name performs significantly better in slow network conditions, where it
delivers $5.9\%$ higher bitrate with $2.4\%$ fewer stalls on average.
When the network connectivity is unstable, ABR is challenging\,---\,a controller must agilely switch
to lower bitrate when the bandwidth prediction or buffer level is low, but must avoid being too conservative by persistently sticking with low bitrates (when is is feasible to use 
higher bitrate without stalling).
In the slow network condition, \name empirically uses the noisy network bandwith estimation better than the heuristic system in order to maintain better buffer levels.
%
This indicates that \name optimizes algorithm performance under network conditions that existing schemes may overlook.
\section{Discussion}
\label{s:discussion}
We intend to work on several directions to further enhance \name in the production systems.
First, \name's training is only performed once offline with pre-collected network traces.
To better incorporate with the updates in the backend infrastructure, we can set up a
continual retraining routine weekly or daily. Prior studies have shown the benefit of
continual training with ever updating systems~\cite{puffer}. 

Second, we primarily evaluate \name on Facebook's web-based video platform, because
it has the fastest codebase update cycle (unlike mobile development, where the updates
are batched in new version releases).
However, the network conditions for cellular networks have larger variability and are more
unpredictable, where the gain of an RL-based ABR scheme can be larger (e.g., we observed
larger performance gain for \name when the network condition is poor in \S\ref{s:analysis}.
Developing a similar learning framework for mobile clients can potentially lead to
larger ABR improvements.

Third, the gains from using \name are rather modest, as they use only the same state variables (\S\ref{s:rl}) as the current heuristic-based ABR algorithm.  Given a fixed parameterization of a simple policy, other techniques such as Bayesian optimization currently serve as a more practical alternative to RL.  However, \name can also  we extend the state space to incorporate more contextual features, such as video streaming regions, temporal information, and the contents of the video itself (since categorizing and optimizing the video quality based on video content types can likely result in better perceptual quality), which engineers cannot easily fold into heuristics. We expect that RL methods provide more practical benefit when the state features become richer. 

%

Lastly, there exists a discrepancy between simulated buffer dynamics and the real video streaming session
in practice. Better bridging this gap can increase the generalizability of \name's learned policy.
To this end, there is ongoing work addressing the discrepancy between simulation and reality
with Bayesian optimization in reward shaping~\cite{letham2019bayesian}. Furthermore, another viable approach is to
directly perform RL training on the production system. The challenge for this is to construct
a similarly safe training mechanism~\cite{shield_rl} that prevents the initial RL trials from decreasing perceptual quality of a video
(e.g., restricting the initial RL policy from randomly select poor bitrates).

\section{Conclusion}
\label{s:conclusion}
We presented \name, a system that uses RL to automatically learn high-quality ABR algorithms for
Facebook's production web-based video platform.
\name has several customized components for solving the challenges in production deployment,
including a scalable architecture for videos with arbitrary bitrates, a variance reduction
RL training method and a Bayesian optimization scheme for reward shaping.
For deployment, we translate \name's policy to an interpretable form for better maintenance and safety.
In a week-long worldwide deployment with more than 30 million video streaming sessions,
our RL approach outperforms the existing carefully-tuned ABR algorithm by at least
$1.6\%$ in video quality and $0.4\%$ in stall.

\bibliography{paper}
\bibliographystyle{icml2019}


\end{document}